\documentclass{sig-alternate}

\makeatletter
\newif\if@restonecol
\makeatother

\usepackage{url}
\usepackage{hyperref}
\usepackage[boxed]{algorithm2e}

\begin{document}
%
\conferenceinfo{SC11}{ November 12-18, 2011, Seattle, Washington, USA}
\CopyrightYear{2011} 
\crdata{978-1-4503-0771-0/11/11}  

\title{Scaling Lattice QCD beyond 100 GPUs}

%
%
%
%
%

\numberofauthors{6} 
%
\author{
%
%
\alignauthor
R. Babich\titlenote{These authors contributed equally to this work.}\\
\affaddr{Center for Computational Science}\\
\affaddr{Boston University}\\
\affaddr{Boston, MA 02215, USA}\\
\email{rbabich@bu.edu}
\alignauthor
M. A. Clark \raisebox{9pt}{$\ast$}\\
       \affaddr{Harvard-Smithsonian Center for Astrophysics}\\
       \affaddr{60 Garden St}\\
       \affaddr{Cambridge, MA 02143, USA}\\
       \email{mikec@seas.harvard.edu}
\alignauthor 
B. Jo\'o \raisebox{9pt}{$\ast$}\\
       \affaddr{Thomas Jefferson National Accelerator Facility}\\
       \affaddr{Newport News, VA 23606, USA}\\
       \email{bjoo@jlab.org}
\and  
\alignauthor G. Shi \raisebox{9pt}{$\ast$}\\
       \affaddr{National Center for Supercomputing Applications}\\
       \affaddr{University of Illinois}\\
       \affaddr{Urbana, IL 61801}\\
       \email{gshi@ncsa.uiuc.edu}
\alignauthor R. C. Brower\\
\affaddr{Center for Computational Science}\\
\affaddr{Boston University}\\
\affaddr{Boston, MA 02215, USA}\\
\email{brower@bu.edu}
\alignauthor S. Gottlieb\\
       \affaddr{Department of Physics}\\
       \affaddr{Indiana University}\\
      \affaddr{Bloomington, IN 47405, USA}\\
      \email{sg@indiana.edu}
}
\date{30 July 1999}

\maketitle
\begin{abstract}
 Over the past five years, graphics processing units (GPUs) have had
  a transformational effect on numerical lattice quantum
  chromodynamics (LQCD) calculations in nuclear and particle
  physics. While GPUs have been applied with great success to the
  post-Monte Carlo ``analysis'' phase which accounts for a substantial
  fraction of the  workload in a typical LQCD calculation, the initial Monte Carlo
  ``gauge field generation'' phase requires capability-level
  supercomputing, corresponding to O(100) GPUs or more. Such strong
  scaling has not been previously achieved. In this contribution, we
  demonstrate that using a multi-dimensional parallelization strategy
  and a domain-decomposed preconditioner allows us to scale into this
  regime. We present results for two popular discretizations of the
  Dirac operator, Wilson-clover and improved staggered, employing up to
  256 GPUs on the Edge cluster at Lawrence Livermore National Laboratory.
\end{abstract}



\bigskip
\section{Introduction}
Lattice QCD (LQCD) is one of the original computational grand
challenges~\cite{grandChallenge}.  Increasingly accurate numerical
solutions of this quantum field theory are being used in tandem with
experiment and observation to gain a deeper quantitative understanding
for a range of phenomena in nuclear and high energy physics.
Advances during the last quarter century required prodigious
computational power, the development of sophisticated algorithms, and
highly optimized software.  As a consequence LQCD is one of the driver
applications that have stimulated the evolution of new architectures
such as the BlueGene series~\cite{QCDOCtoBG}.  Graphics processing
unit (GPU) clusters challenge us to adapt lattice field theory
software and algorithms to exploit this potentially transformative
technology. Here we present methods allowing QCD linear solvers to
scale to hundreds of GPUs with high efficiency.  The resulting
multi-teraflop performance is now comparable to typical QCD codes
running on capability machines such as the Cray and the BlueGene/P
using several thousand cores.

GPU computing has enjoyed a rapid growth in popularity in recent years
due to the impressive performance to price ratio offered by the
hardware and the availability of free software development tools and
example codes.  Currently, the fastest supercomputer in the world,
Tianhe-1A, is a GPU-based system, and several large-scale GPU
systems are either under consideration or are in active development.
Examples include the Titan system proposed for the Oak Ridge Leadership
Computing Facility (OLCF) and the NSF Track 2 Keeneland system to be
housed at the National Institute for Computational Sciences (NICS).

Such systems represent a larger trend toward heterogeneous
architectures, characterized not only by multiple processor types (GPU
and conventional CPU) with very different capabilities, but also by a
deep memory hierarchy exhibiting a large range of relevant bandwidths
and latencies.  These features are expected to typify at least one
path (or ``swim-lane'') toward exascale computing.  The intrinsic
imbalance between different subsystems can present bottlenecks and a
real challenge to application developers.  In particular, the PCI-E
interface currently used to connect CPU, GPU, and communications fabric
on commodity clusters can prove a severe impediment for strong-scaling
the performance of closely coupled, nearest-neighbor stencil-like
codes into the capability computing regime. Overcoming such
limitations is vital for applications to succeed on future large-scale
heterogeneous resources.

We consider the challenge of scaling LQCD codes to a large number of
GPUs.  LQCD is important in high energy and nuclear physics as it is
the only currently known first-principles non-perturbative approach for calculations
involving the strong force. Not only is LQCD important from the point
of view of physics research, but historically LQCD codes have often
been used to benchmark and test large scale computers. The balanced
nature of QCD calculations, which require approximately 1 byte/flop
in single precision, as well as their regular memory access and nearest
neighbor communication patterns have meant that LQCD codes could be
deployed quickly, scaled to large partitions, and used to exercise the
CPU, memory system, and communications fabric. In GPU computing, LQCD
has been highly successful in using various forms of data compression and mixed
precision solvers \cite{Clark:2009wm} to alleviate memory bandwidth
contraints on the GPU in order to attain high performance.  Our
multi-GPU codes
\cite{Babich:2010:PQL:1884643.1884695,Gottlieb:2010zz,Shi:2011ipdps}
are in production, performing capacity analysis calculations on systems at several
facilities, including Lincoln and EcoG (NCSA), Longhorn (TACC), the ``9g''
and ``10g'' clusters at Jefferson Laboratory (JLab), and Edge (LLNL).

The challenge is now to scale LQCD computations into the O(100) GPU
regime, which is required if large GPU systems are to replace the more
traditional massively parallel multi-core supercomputers that are
used for the gauge generation step of LQCD.  Here, capability-class
computing is required since the algorithms employed consist of single
streams of Monte Carlo Markov chains, and so require strong scaling.

Our previous multi-GPU implementations utilized a strategy of
parallelizing in the time direction only, using traditional
iterative Krylov solvers (e.g., conjugate gradients).  This severely
limits the number of GPUs that can be used and thus maximum
performance.  In order to make headway, it has become important to
parallelize in additional dimensions in order to give sublattices with
improved surface-to-volume ratios and to explore algorithms that
reduce the amount of communication altogether such as domain
decomposed approaches.

In this paper, we make the following contributions: (i) we parallelize
the application of the discretized Dirac operator in the QUDA library
to communicate in multiple dimensions, (ii) we investigate the utility
of an additive Schwarz domain-decomposed preconditioner for the
Generalized Conjugate Residual (GCR) solver, and (iii) we perform
performance tests using (i) and (ii) on partitions of up to 256 GPUs.
The paper is organized as follows: we present an outline of LQCD
computations in Sec. \ref{s:LQCD} and discuss iterative linear solvers
in Sec. \ref{s:Linear}.  A brief overview of previous and related work
is given in Sec. \ref{s:Previous}, and in Sec. \ref{s:QUDA} we
describe the QUDA library upon which this work is based.  The
implementation of the multi-dimensional, multi-GPU parallelization is
discussed in Sec. \ref{s:MultiDim}.  The construction of our optimized
linear solvers are elaborated in Sec. \ref{s:Solver}. We present
performance results in Sec. \ref{s:Results}. Finally, we summarize and
conclude in Sec. \ref{s:Conclusions}.

\section{Lattice QCD} \label{s:LQCD} 
Weakly coupled field theories such as quantum electrodynamics
can by handled with perturbation theory.
In QCD, however, at low
energies perturbative expansions fail and a non-perturbative 
method is required. Lattice QCD is the only known, model
independent, non-perturbative tool currently available to perform
QCD calculations. 

LQCD calculations are typically Monte-Carlo evaluations of a 
Euclidean-time path integral.  A sequence of configurations of the gauge fields
is generated in a process known as {\em
  configuration generation}. The gauge configurations are
importance-sampled with respect to the lattice action and
represent a snapshot of the QCD vacuum.  Configuration generation is
inherently sequential as one configuration is generated from the
previous one using a stochastic evolution process.  Many variables can
be updated in parallel and the focused power of capability computing
systems has been essential.  Once the field configurations have been
generated, one moves on to the second stage of the calculation, known
as {\em analysis}. In this phase, observables of interest are
evaluated on the gauge configurations in the ensemble, and the results
are then averaged appropriately, to form {\em ensemble averaged}
quantities. It is from the latter that physical results such as
particle energy spectra can be extracted.  The analysis phase can be
task parallelized over the available configurations in an ensemble and
is thus extremely suitable for capacity level work on clusters, or
smaller partitions of supercomputers.

\subsection{Dirac PDE discretization}

The fundamental interactions of QCD, those taking place between quarks
and gluons, are encoded in the quark-gluon interaction differential
operator known as the Dirac operator.  A proper discretization of the
Dirac operator for lattice QCD requires special care.  As is common in PDE
solvers, the derivatives are replaced by finite differences.  Thus on
the lattice, the Dirac operator becomes a large sparse matrix, \(M\),
and the calculation of quark physics is essentially reduced to many
solutions to systems of linear equations given by
\begin{equation}
Mx = b.
\label{eq:linear}
\end{equation}
Computationally, the brunt of the numerical work in LQCD for both the
gauge generation and analysis phases involves solving such linear
sytems.

A small handful of discretizations are in common use, differing in
their theoretical properties.  Here we focus on two of the most widely-used forms
for $M$, the Sheikholeslami-Wohlert
\cite{Sheikholeslami:1985ij} form (colloquially known as {\em
  Wilson-clover}), and the improved staggered form, specifically the 
$a^2$ tadpole-improved (\emph{asqtad}) formulation \cite{RevModPhys.82.1349}.

\subsection{Wilson-clover matrix}
The Wilson-clover matrix is a central-difference discretization of the
Dirac operator, with the addition of a diagonally-scaled Laplacian to
remove the infamous fermion doublers (which arise due to the red-black
instability of the central-difference approximation).  When acting in
a vector space that is the tensor product of a 4-dimensional
discretized Euclidean spacetime, {\it spin} space, and {\it color}
space it is given by
\begin{align}
 M_{x,x'}^{WC} &= - \frac{1}{2} \displaystyle \sum_{\mu=1}^{4} \bigl(
 P^{-\mu} \otimes U_x^\mu\, \delta_{x+\hat\mu,x'}\, + P^{+\mu} \otimes
 U_{x-\hat\mu}^{\mu \dagger}\, \delta_{x-\hat\mu,x'}\bigr) \nonumber \\
&\quad\, + (4 + m + A_x)\delta_{x,x'} \nonumber \\
&\equiv  - \frac{1}{2}D_{x,x'}^{WC} + (4 + m + A_{x}) \delta_{x,x'}.
\label{eq:Mclover}
\end{align} 
Here \(\delta_{x,y}\) is the Kronecker delta; \(P^{\pm\mu}\) are
\(4\times 4\) matrix projectors in {\it spin} space; \(U\) is the QCD
gauge field which is a field of special unitary $3\times 3$ (i.e.,
SU(3)) matrices acting in {\it color} space that live between the
spacetime sites (and hence are referred to as link matrices); \(A_x\)
is the \(12\times12\) clover matrix field acting in both spin and
color space,\footnote{Each clover matrix has a Hermitian block
  diagonal, anti-Hermitian block off-diagonal structure, and can be
  fully described by 72 real numbers.} corresponding to a first order
discretization correction; and \(m\) is the quark mass parameter.  The
indices \(x\) and \(x'\) are spacetime indices (the spin and color
indices have been suppressed for brevity).  This matrix acts on a
vector consisting of a complex-valued 12-component \emph{color-spinor}
(or just {\em spinor}) for each point in spacetime.  We refer to the
complete lattice vector as a spinor field.

\begin{figure}[htb]
\begin{center}
\includegraphics[width=2.5in]{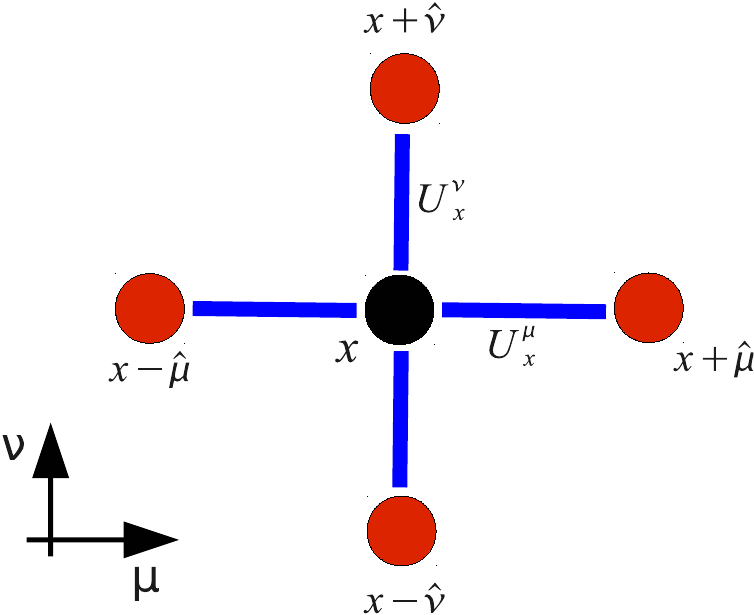}
\end{center}
\caption{\label{fig:dslash}The nearest neighbor stencil part of the
 lattice Dirac operator $D$, as defined in (\ref{eq:Mclover}), in the $\mu-\nu$
 plane.  The \emph{color-spinor} fields are located on the
 sites. The SU(3) color matrices $U^\mu_x$ are associated with the links. The
 nearest neighbor nature of the stencil suggests a natural even-odd (red-black)
 coloring for the sites.}
\end{figure}

\subsection{Improved staggered matrix}

The staggered matrix again is a central-difference discretization of
the Dirac operator; however, the fermion doublers are removed
through ``staggering'' the spin degrees of freedom onto neighboring
lattice sites.  This essentially reduces the number of spin degrees of
freedom per site from four to one, which reduces the computational
burden significantly.  This transformation, however, comes at the
expense of increased discretization errors, and breaks the so-called
quark-flavor symmtery.  To reduce these discretization errors, the
gauge field that connects nearest neighboring sites on the lattice
(\(U^\mu_x\) in Equation \ref{eq:Mclover}) is smeared, which essentially is
a local averaging of the field.  There are many prescriptions for this
averaging; and here we employ the popular asqtad 
procedure \cite{RevModPhys.82.1349}.  The errors are
further reduced through the inclusion of the third neighboring spinors
in the derivative approximaton.  The asqtad matrix is given by
\begin{align}
  M_{x,x'}^{IS} &= - \frac{1}{2} \displaystyle \sum_{\mu=1}^{4} \bigl(
  \hat{U}_x^\mu\, \delta_{x+\hat\mu,x'}\, + \hat{U}_{x-\hat\mu}^{\mu \dagger}\, \delta_{x-\hat\mu,x'} +\nonumber\\
&  \quad\quad\quad\quad \check{U}_x^\mu\, \delta_{x+3\hat\mu,x'}\, + \check{U}_{x-3\hat\mu}^{\mu \dagger}\, \delta_{x-3\hat\mu,x'}\bigr) + m\delta_{x,x'} \nonumber \\
  &\equiv - \frac{1}{2}D_{x,x'}^{IS} + m\delta_{x,x'}.
\label{eq:Mstag}
\end{align} 
Unlike \(M^{WC}\), the matrix \(M^{IS}\) consists solely of a
derivative term \(D^{IS}\) and the mass term.  There are two gauge
fields present: \(\hat{U}^\mu_x\) is the {\it fat} gauge field, and is
the field produced from locally averaging \(U^\mu_x\); and
\(\check{U}^\mu_x\) is the {\it long} gauge field produced by taking
the product of the links \(U^\mu_x U^\mu_{x+\hat{\mu}}
U^\mu_{x+2\hat{\mu}}\).  While both of these fields are functions of
the original field \(U^\mu_x\), in practice, these fields are
pre-calculated before the application of \(M^{IS}_{x,x'}\) since
iterative solvers will require the application of \(M^{IS}_{x,x'}\)
many hundreds or thousands of times.  Since there are no separate spin
degrees of freedom at each site, this matrix acts on a vector of
complex-valued 3-component colors; however, for convenience we nevertheless
refer to the complex lattice vector as a spinor field.

\section{Iterative solvers}
\label{s:Linear}

\subsection{Krylov solvers}
For both discretizations under consideration, \(M\) is a large sparse matrix, and iterative Krylov
solvers are typically used to obtain solutions to Equation (\ref{eq:linear}),
requiring many repeated evaluations of the sparse matrix-vector
product.  The Wilson-clover matrix is non-Hermitian, so either
Conjugate Gradients \cite{Hestenes:1952} on the normal equations (CGNE
or CGNR) is used, or more commonly, the system is solved directly
using a non-symmetric method, e.g., BiCGstab \cite{vanDerVorst:1992}.
Even-odd (also known as red-black) preconditioning is almost always
used to accelerate the solution finding process for this system, where
the nearest neighbor property of the \(D^{WC}\) matrix is exploited
to solve the Schur complement system~\cite{Degrand1990211}. 

The staggered fermion matrix is anti-Hermitian, and has the
convenient property that when multiplied by its Hermitian conjugate,
the even and odd lattices are decoupled and can be solved independently
from each other.  There are no commonly used preconditioners for the
staggered matrix.  When simulating asqtad fermions, for both the gauge
field generation and for the analysis stages, one is confronted with
solving problems of the form
\begin{equation}
(M^\dagger M + \sigma_i I) x_i = b\qquad{i=1\ldots N}
\label{eq:multishift}
\end{equation}
where \(\sigma_i\) is a constant scalar and \(I\) is the
identity matrix.  This is equivalent to solving \(N\) different linear
systems at different mass parameters for a constant source \(b\).
Since the generated Krylov spaces are the same for each of these
linear systems, one can use a multi-shift solver (also known as a
multi-mass solver) to produce all \(N\) solutions simultaneously in
the same number of iterations as the smallest shift (least well
conditioned)~\cite{Jegerlehner:1996pm}.

In both cases, the quark mass controls the condition number of the
matrix, and hence the convergence of such iterative solvers.
Unfortunately, physical quark masses correspond to nearly indefinite
matrices.  Given that current lattice volumes are at least \(10^8\)
degrees of freedom in total, this represents an extremely
computationally demanding task.  For both the gauge generation and
analysis stages, the linear solver accounts for 80--99\% of the
execution time.

\subsection{Additive Schwarz preconditioner}

As one scales lattice calculations to large core counts, on
leadership-class partitions, one is faced with a strong-scaling
challenge: as core counts increase, for a fixed global lattice volume
the local sub-volume per core decreases and the surface-to-volume ratio
of the local sub--lattice increases. Hence, the ratio of communication
to local computation also grows. For sufficiently many cores, it
becomes impossible to hide communication by local computation and the
problem becomes communications bound, at the mercy of the system
interconnect. This fact, with the need for periodic global reduction
operations in the Krylov solvers, leads to a slowdown of their
performance in the large scaling limit. For GPU-based clusters, where
inter-GPU communication is gated by the PCI-E bus, this limitation is
substantially more pronounced and can occur in partitions as small as
O(10) GPUs or less.

This challenge of strong scaling of traditional Krylov solvers
motivates the use of solvers which minimize the amount of
communication.  Such solvers are commonly known as
domain-decomposition solvers and two forms of them are commonly used:
multiplicative Schwarz and additive Schwarz processes~\cite{schwarz}.
In this work we focus upon the additive Schwarz method.  Here, the
entire domain is partitioned into blocks which may or may not overlap.
The system matrix is then solved within these blocks, imposing
Dirichlet (zero) boundary conditions at the block boundaries.  The
imposition of Dirichlet boundary conditions means that no
communication is required between the blocks, and that each block can
be solved independently.  It is therefore typical to assign the blocks
to match the sub-domain assigned to each processor in a parallel
decomposition of a domain.  A tunable parameter in these solvers is
the degree of overlap of the blocks, with a greater degree of overlap
corresponding to increasing the size of the blocks, and hence the
amount of computation required to solve each block.  A larger overlap
will typically lead to requiring fewer iterations to reach
convergence, since, heuristically, the larger sub blocks, will
approximate better the original matrix and hence their inverses will
form better preconditioners.  Note that an additive Schwarz solver
with non-overlapping blocks is equivalent to a block-Jacobi solver.

Typically, Schwarz solvers are not used as a standalone solver, but rather
they are employed as preconditioners for an outer Krylov method.
Since each local system matrix is usually solved using an iterative
solver, this requires that the outer solver be a {\em flexible} solver.
Generalized conjugate residual (GCR) is such a solver, and
we shall employ it for the work in this paper.

\section{Overview of related work} \label{s:Previous}

Lattice QCD calculations on GPUs were originally reported in
\cite{Egri2007631} where the immaturity of using GPUs for general
purpose computation necessitated the use of graphics APIs.  Since the
advent of CUDA in 2007, there has been rapid uptake by the LQCD
community (see \cite{Clark:2009pk} for an overview).  More recent work
includes \cite{Alexandru:2011ee}, which targets the computation of
multiple systems of equations with Wilson fermions where the systems
of equations are related by a linear shift.  Solving such systems is
of great utility in implementing the overlap formulation of QCD.  This
is a problem we target in the staggered-fermion solver below.  The
work in \cite{Chiu:2011rc} targets the domain-wall fermion formulation
of LQCD.  This work concerns the QUDA library \cite{Clark:2009wm},
which we describe in Sec.\ \ref{s:QUDA} below.

Most work to date has concerned single-GPU LQCD implementations, and
beyond the multi-GPU parallelization of
QUDA~\cite{Babich:2010:PQL:1884643.1884695,Shi:2011ipdps} and the
work in \cite{Alexandru:2011sc} which targets a multi-GPU
implementation of the overlap formulation, there has been little
reported in the literature, though we are aware of other
implementations which are in production~\cite{Borsani}.
 
Domain-decomposition algorithms were first introduced to LQCD in
\cite{Luscher:2003qa}, through an implementation of the Schwarz
Alternating Procedure preconditioner, which is a multiplicative
Schwarz preconditioner. More akin to the work presented here is the
work in \cite{Osaki:2010vj} where a restricted additive Schwarz
preconditioner was implemented for a GPU cluster. However, the work
reported in \cite{Osaki:2010vj} was carried out on a rather small
cluster containing only 4 nodes and connected with Gigabit Ethernet.
The work presented here aims for scaling to O(100) GPUs using a QDR
Infiniband interconnect.

\section{QUDA}\label{s:QUDA}
The QUDA library is a package of optimized CUDA kernels and wrapper
code for the most time-consuming components of an LQCD computation.
It has been designed to be easy to interface to existing code bases,
and in this work we exploit this interface to use the popular LQCD
applications Chroma and MILC.  The QUDA library has attracted a
diverse developer community of late and is being used in production at
LLNL, Jlab and other U.S.\ national laboratories, as well as in Europe.
The latest development version is always available in a publically-accessible
source code repository~\cite{githubQUDA}.

QUDA implements optimized linear solvers, which when running on a
single GPU achieve up to 24\% of the GPU's peak performance through
aggressive optimization.  The general strategy is to assign a single
GPU thread to each lattice site, each thread is then responsible for
all memory traffic and operations required to update that site on the lattice
given the stencil operator.  Maximum memory bandwidth is obtained
by reordering the spinor and gauge fields to achieve memory coalescing
using structures of float2 or float4 arrays, and using the texture
cache where appropriate.  Memory traffic reduction is employed where
possible to overcome the relatively low arithmetic intensity of the
Dirac matrix-vector operations, which would otherwise limit
performance.  Strategies include: (a) using compression for the
\(SU(3)\) gauge matrices to reduce the 18 real numbers to 12 (or 8)
real numbers at the expense of extra computation; (b) using similarity
transforms to increase the sparsity of the Dirac matrices; (c)
utilizing a custom 16-bit fixed-point storage format (here on referred
to as half precision) together with mixed-precision linear solvers to
achieve high speed with no loss in accuracy. 

Other important computational kernels provided in the library include
gauge field smearing routines for constructing the fat gauge field used
in the asqtad variant of the improved staggered discretization, as well as
force term computations required for gauge field generation.

The extension of QUDA to support multiple GPUs was reported in
\cite{Babich:2010:PQL:1884643.1884695}, where both strong and weak
scaling was performed on up to 32 GPUs using a lattice volume of
\(32^3\times256\) with Wilson-clover fermions.  This employed
partitioning of the lattice along the time dimension only, and was
motivated by expediency and the highly asymmetric nature of
the lattices being studied.  While this strategy was sufficient to achieve excellent
(artificial) weak scaling performance, it severely limits the strong
scaling achievable for realistic volumes because of the increase in
surface-to-volume ratio.  The application of these strategies to the
improved staggered discretization was described in \cite{Shi:2011ipdps},
where strong scaling was achieved on up to 8 GPUs using a lattice
volume of \(24^3\times96\).  Here the single dimensional
parallelization employed restricts scaling more severely than for
Wilson-clover because of the 3-hop stencil of the improved staggered
operator which decreases the locality of the operator.


\section{Multi-dimensional partitioning} \label{s:MultiDim}

\subsection{General strategy}

Because lattice discretizations of the Dirac operator generally only couple
sites of the lattice that are nearby in spacetime, the first step in any
parallelization strategy is to partition the lattice.
As indicated above, prior to this work multi-GPU parallelization of
the QUDA library had been carried out with the lattice partitioned
along only a single dimension.  The time ($T$) dimension was
chosen, first because typical lattice volumes are asymmetric with $T$
longest, and secondly because this dimension corresponds to the
slowest varying index in memory in our implementation, making it
possible to transfer the boundary face from GPU to host with a
straightforward series of memory copies.  Going beyond this approach
requires much more general handling of data movement, computation,
and synchronization, as we explore here.

In the general case, upon partitioning the lattice each GPU is
assigned a 4-dimensional subvolume that is bounded by at most eight
3-dimensional ``faces.''  Updating sites of the spinor field on or
near the boundary requires data from neighboring GPUs.  The data
received from a given neighbor is stored in a dedicated buffer on the
GPU which we will refer to as a ``ghost zone'' (since it shadows data
residing on the neighbor).  Computational kernels are modified so as
to be aware of the partitioning and read data from the appropriate
location --- either from the array corresponding to the local subvolume
or one of the ghost zones.  Significant attention is paid to
maintaining memory coalescing and avoiding thread divergence, as
detailed below.

\begin{figure}[htb]
\begin{center}
\includegraphics[width=3.5in]{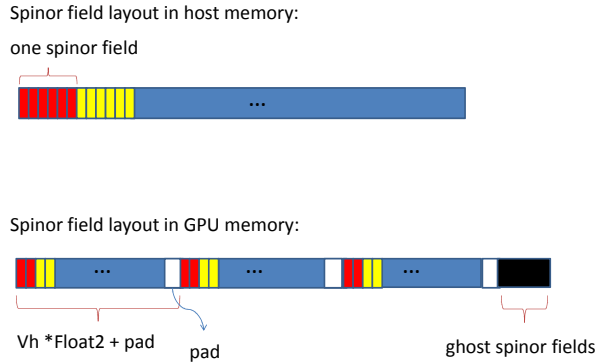}
\end{center}
\caption{\label{fig:spinor_layout} Spinor field layout in host and GPU
  memory for the staggered discretization (consisting of 6 floating
  point numbers per site).  Here Vh is half the local volume
  of the lattice, corresponding to the number of sites in an even/odd
  subset.  Layout for the Wilson-clover discretization is similar,
  wherein the spinor field consists of 24 floating point numbers per
  site.}
\end{figure}

\begin{figure}[htb]
\begin{center}
\includegraphics[width=3.5in]{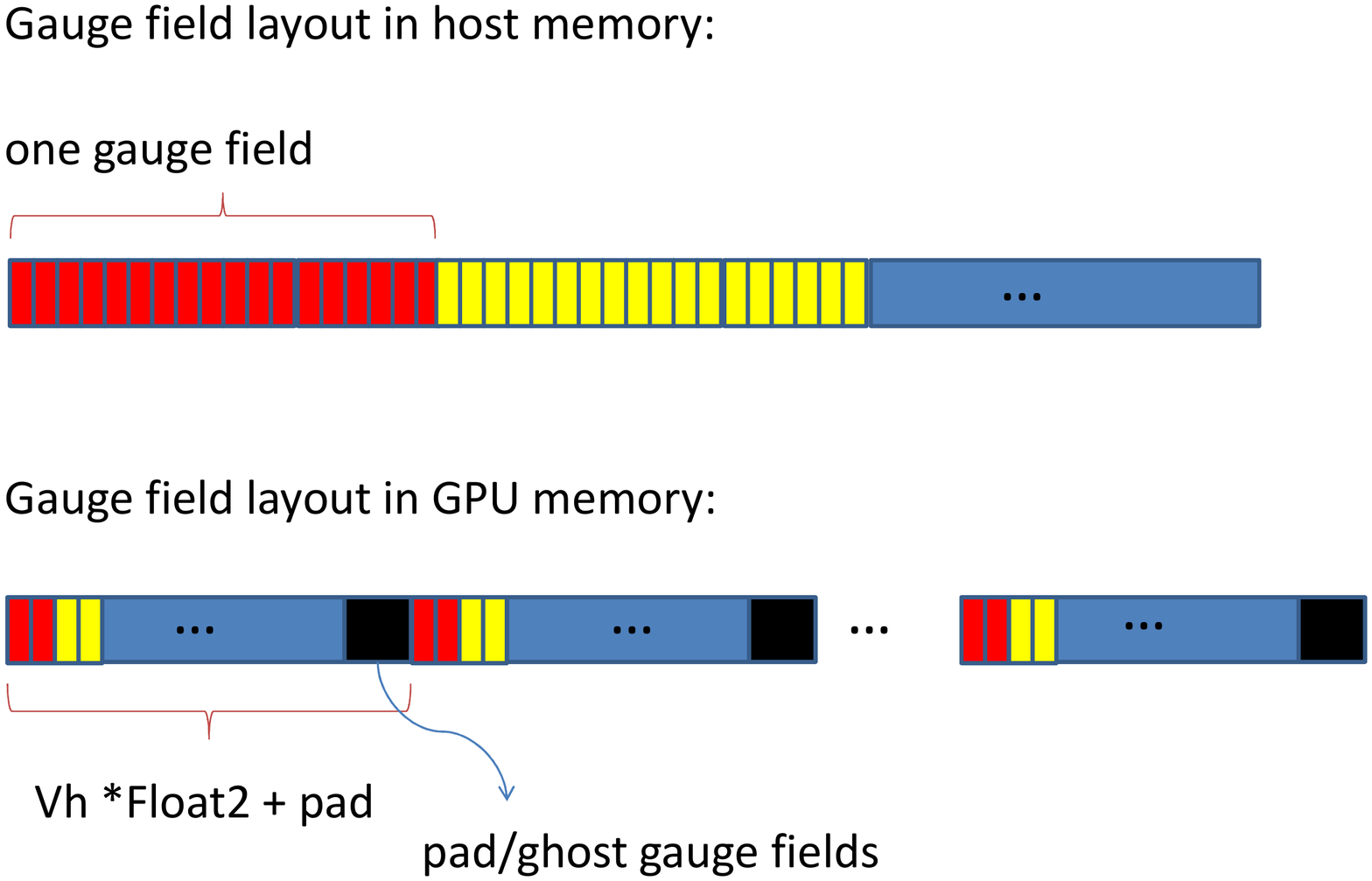}
\end{center}
\caption{\label{fig:gauge_layout}Gauge field layout in host and GPU
  memory.  The gauge field consists of 18 floating point numbers per
  site (when no reconstruction is employed) and is ordered on the GPU
  so as to ensure that memory accesses in both interior and
  boundary-update kernels are coalesced to the extent possible.}
\end{figure}

Ghost zones for the spinor field are placed in memory after the local
spinor field so that BLAS-like routines, including global reductions,
may be carried out efficently.  While the ghost spinor data for the T
dimension is contiguous and can be copied without a gather operation,
the ghost spinor data for the other three dimensions must be collected
into contiguous GPU memory buffers by a GPU kernel before it can be
transfered to host memory.  The ghost zone buffers are then exchanged
between neighboring GPUs (possibly residing in different nodes). Once
inter-GPU communication is complete, the ghost zones are copied to
locations adjoining the local array in GPU memory.  Allocation of
ghost zones and data exchange in a given dimension only takes place
when that dimension is partitioned, so as to ensure that GPU memory as well as
PCI-E and interconnect bandwidth are not wasted.  Layout of the local
spinor field, ghost zones, and padding regions are shown in
Fig.~\ref{fig:spinor_layout}.  The padding region is of adjustable
length and serves to reduce partition
camping~\cite{Clark:2009wm,Ruetsch:2009pc} on previous-generation
NVIDIA GPUs.\footnote{This is less a concern for the Tesla M2050 cards
  used in this study, as the Fermi memory controller employs
  address hashing to alleviate this problem.}  The gauge field is
allocated with a similar padding region, and we use this space to
store ghost zones for the gauge field, which must only be transfered
once at the beginning of a solve.  The layout is illustrated in
Fig.~\ref{fig:gauge_layout}.

For communication, our implementation is capable of employing either
of two message-passing frameworks -- MPI or QMP.  The latter ``QCD
message-passing'' standard was originally developed to provide a
simplified subset of communication primitives most used by LQCD codes,
allowing for optimized implementations on a variety of architectures,
including purpose-built machines that lack MPI.  Here we rely on the
reference implementation, which serves as a thin layer over MPI itself
(but nevertheless serves a purpose as the communications interface used
natively by Chroma, for example).  Accordingly, performance with the
two frameworks is virtually identical.  At present, we assign GPUs to
separate processes which communicate via message-passing.  Exploration
of peer-to-peer memory copies, recently added in CUDA 4.0, and host-side
multi-threading is underway.

\subsection{Interior and exterior kernels}

In~\cite{Babich:2010:PQL:1884643.1884695,Shi:2011ipdps}, where only
the time dimension of the lattice was partitioned, we separated the
application of the Dirac operator into two kernels, one to update
sites on the boundaries of the local sublattice (the \emph{exterior
  kernel}) and one to perform all remaining work (the \emph{interior
  kernel}).  Here we extend this approach by introducing one exterior
kernel for every dimension partitioned, giving a total of four
exterior kernels in the most general case.  The interior kernel
executes first and computes the spinors interior to the subvolume, as
well as any contributions to spinors on the boundaries that do not
require data from the ghost zones.  For example, if a spinor is
located only on the T+ boundary, the interior kernel computes
contributions to this spinor from all spatial dimensions, as well as
that of the negative T direction.  The contribution from the positive
T direction will be computed in the T exterior kernel using the ghost
spinor and gauge field from the T+ neighbor.  Since spinors on the
corners belong to multiple boundaries, they receive contributions from
multiple exterior kernels.  This introduces a data dependency between
exterior kernels, which must therefore be executed sequentially.

Another consideration is the ordering used for assigning threads to
sites.  For the interior kernel and T exterior kernel, the
one-dimensional thread index (given in CUDA C by {\tt
  (blockIdx.x*blockDim.x + threadIdx.x)}) is assigned to sites of the
four-dimensional sublattice in the same way that the spinor and gauge
field data is ordered in memory, with X being the fastest varying
index and T the slowest.  It is thus guaranteed that all spinor and
gauge field accesses are coalesced.  In the X,Y,Z exterior kernels,
however, only the destination spinors are indexed in this way, while
the ghost spinor and gauge field are indexed according to a different
mapping.  This makes it impossible to guarantee coalescing for both
reads and writes; one must choose one order or the other for assigning
the thread index.  We choose to employ the standard T-slowest mapping
for the X,Y,Z exterior kernels to minimize the penalty of uncoalesced
accesses, since a greater fraction of the data traffic comes from the
gauge field and source spinors.

\subsection{Computation, communication, and streams}

\begin{figure}[htb]
\begin{center}
\includegraphics[width=3.5in]{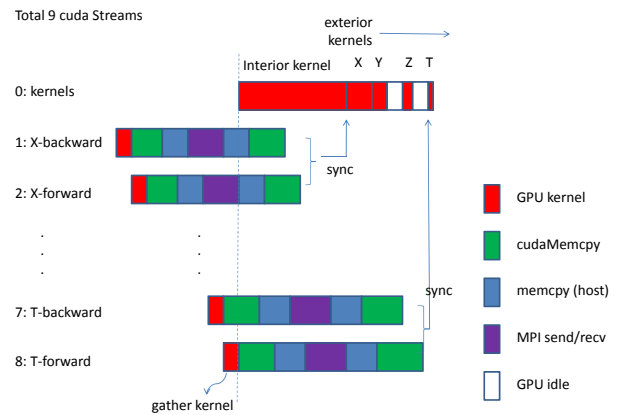}
\end{center}
\caption{\label{fig:overlap}Usage of CUDA streams in the application
  of the Dirac operator, illustrating the multiple stages of
  communication.  A single stream is used for the interior and
  exterior kernels, and two streams per dimension are used for gather
  kernels, PCI-E data transfer, host memory copies, and inter-node
  communication.}
\end{figure}

Our implementation employs CUDA streams to overlap computation with
communication, as well as to overlap GPU-to-host with inter-node
communication.  Two streams per dimension are used, one for gathering
and exchanging spinors in the forward direction and the other in the
backward direction.  One additional stream is used for executing the
interior and exterior kernels, giving a total of 9 streams as shown in
Fig.\ \ref{fig:overlap}.  The gather kernels for all dimensions are
launched on the GPU immediately so that communication in all
directions can begin.  The interior kernel is executed after all
gather kernels finish, overlapping completely with the communication.
We use different streams for different dimensions so that the
different communication components can overlap with each other,
including the device-to-host memory copy, the copy from pinned host
memory to pagable host memory, the MPI send and receive, the memory
copy from pagable memory to pinned memory on the receiving side, and
the final host-to-device memory copy.  While the interior kernel can
be overlapped with communications, the exterior kernels must wait for
arrival of the ghost data.  As a result, the interior kernel and
subsequent exterior kernels are placed in the same stream, and each
exterior kernel blocks waiting for communication in the corresponding
dimension to finish.  For small subvolumes, the total communication
time over all dimensions is likely to exceed the interior kernel run
time, resulting in some interval when the GPU is idle (see
Fig.\ \ref{fig:overlap}) and thus degrading overall performance.

When communicating over multiple dimensions with small subvolumes, the
communication cost dominates over computation, and so any reduction in
the communication is likely to improve performance. The two host
memory copies are required due to the fact that GPU pinned memory is
not compatible with memory pinned by MPI implementations; GPU-Direct
\cite{gpudirect} was not readily available on the cluster used in this
study.  We expect to be able to remove these extra memory copies in
the future when better support from GPU and MPI vendors is
forthcoming.  CUDA 4.0, recently released, includes a promising
GPU-to-GPU direct communication feature that we will explore in the
future to further reduce the communication cost.

\section{Dirac operator performance}

\subsection{Hardware description}

For the numerical experiments discussed in this paper we used the Edge
visualization cluster installed at Lawrence Livermore National
Laboratory. Edge is comprised of a total of 216 nodes, of which 206
are compute nodes available for batch jobs. Each compute node is
comprised of dual-socket six-core Intel X5660 Westmere CPUs running at
2.8GHz and two NVIDIA Tesla M2050 GPUs, running with error correction
(ECC) enabled. The two GPUs share a single x16 PCI-E connection to the
I/O hub (IOH) via a switch.  Eight of the remaining PCI-E lanes serve
a quad data rate (QDR) InfiniBand interface which can thus run at full
bandwidth.  The compute nodes run a locally maintained derivative of a
CentOS 5 kernel with revision {\tt 2.6.18-chaos103}.

To build and run our software we used OpenMPI version 1.5 built on top
of the system GNU C/C++ compiler version 4.1.2. To build and link
against the QUDA  library we used release
candidate 1 of CUDA version 4.0 with driver version 270.27. 

\subsection{Wilson-clover}

\begin{figure}[htb]
\begin{center}
\includegraphics[width=3.5in]{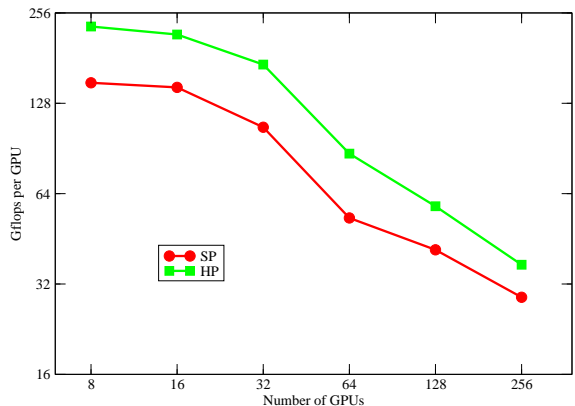}
\end{center}
\caption{\label{fig:clover-dslash}Strong-scaling results for the Wilson-clover
  operator in single (SP) and half (HP) precisions (\(V = 32^3\times256\), 12 gauge reconstruction).}
\end{figure}

In Fig.\ \ref{fig:clover-dslash}, we show the strong scaling of the
Wilson-clover operator on up to 256 GPUs. We see significant departures
from ideal scaling for more than 32 GPUs, as increasing the surface-to-volume
ratio increases the amount of time spent in communication, versus
computation. It seems that, for more than 32
GPUs, we are no longer able to sufficiently overlap computation with 
communication, and the implementation becomes communications
bound.  We note also that as the communications overhead grows, 
the performance advantage of the half precision operator over the 
single precision operator appears diminished. The severity of the
scaling violations seen here highlights the imbalance between the
communications and compute capability of GPU clusters. To overcome
this constraint, algorithms which reduce communication, such 
as the domain-decomposition algorithms described below, are absolutely
essential.

\subsection{Improved staggered}

\begin{figure}[htb]
\begin{center}
\includegraphics[width=3.5in]{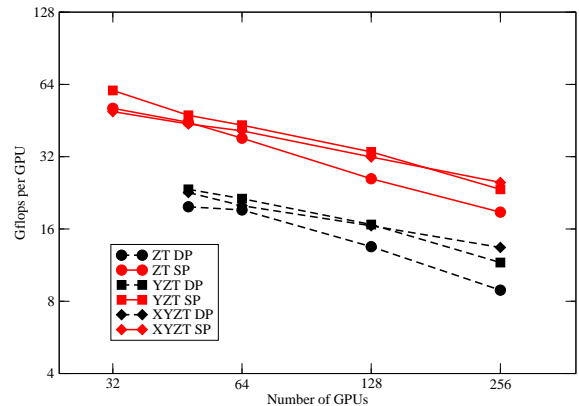}
\end{center}
\caption{\label{fig:asqtad-dslash}Strong-scaling performance 
  for the asqtad operator in double (DP) and single (SP) precision.
  The legend labels denote which dimensions are partitioned between
  GPUs (\(V = 64^3\times192\), no gauge reconstruction).  }
\end{figure}

In Fig.\ \ref{fig:asqtad-dslash}, we plot the performance per GPU for
the asqtad volume used in this study.  A number of interesting
observations can be made about this plot.  At a relatively low number
of GPUs, where we are less communications-bound, having faster kernel
performance is more important than the optimal surface-to-volume
ratio.  As the number of GPUs is increased, the minimization of the
surface-to-volume ratio becomes increasingly important, and the XYZT
partitioning scheme, which has the worst single-GPU performance,
obtains the best performance on 256 GPUs.

\section{Building scalable solvers} 
\label{s:Solver}

\subsection{Wilson-clover additive Schwarz preconditioner}

The poor scaling of the application of the Wilson-clover matrix at a
high number of GPUs at this volume motivates the use of
communication-reducing algorithms.  In this initial work, we
investigate using a non-overlapping additive Schwarz preconditioner
for GCR.  In the text that follows, we refer to this algorithm as
GCR-DD.

Implementation of the preconditioner is simple: essentially, we just
have to switch off the communications between GPUs.  This means that
in applying the Dirac operator, the sites that lie along the
communication boundaries only receive contributions from sites local
to that node.  Additionally, since the solution in each domain is
independent from every other, the reductions required in each of the
domain-specific linear solvers are restricted to that domain only.
As a result the solution of the preconditioner linear system
will operate at similar efficiency to the equivalent single-GPU
performance at this local volume.  The imposition of the Dirichlet
boundary conditions upon the local lattice leads to a vastly reduced
condition number.  This, coupled with the fact that only a very loose
approximation of the local system matrix is required, means that only
a small number of steps of minimum residual (MR) are required to
achieve satisfactory accuracy.

The outer GCR solver builds a Krylov subspace, within which the
residual is minimized and the corresponding solution estimate is
obtained.  Unlike solvers with explicit recurrence relations, e.g.,
CG, the GCR algorithm for the general matrix problem requires explicit
orthogonalization at each step with respect to the previously
generated Krylov space.  Thus, the size of the Krylov space is limited
by the computational and memory costs of orthogonalization.  After the
limit of the Krylov space has been reached, the algorithm is
restarted, and the Krylov space is rebuilt.  

Similar to what was reported for BiCGstab~\cite{Clark:2009wm}, we
have found that using mixed precision provides significant
acceleration of the GCR-DD solver algorithm.  We exclusively use
half precision for solving the preconditioned system.  This is natural
since only a very loose approximation is required.  Additionally, the
restarting mechanism of GCR provides a natural opportunity for using
mixed precision: the Krylov space is built up in low precision and
restarted in high precision.  This approach also conserves the limited
GPU memory, allowing for larger Krylov spaces to be built, albeit in
lower precision.  We follow the implicit solution update scheme
described in \cite{Luscher:2003qa} since this reduces the
orthogonalization overhead.  For the physics of interest, the
inherent noise present in the Monte Carlo gauge generation process is such
that single-precision accuracy is sufficient.  Thus we have found best
performance using a single-half-half solver, where the GCR restarting
is performed in single precision, the Krylov space construction and
accompanying orthogonalization is done in half precision, and
the preconditioner is solved in half precision.  In minimizing
the residual in half precision, there is the inherent risk of the iterated
residual straying too far from the true residual.  Thus, we have added
an early termination criteria for the Krylov subspace generation,
where if the residual is decreased by more than a given tolerance
\(\delta\) from the start of the Krylov subspace generation, the
algorithm is restarted.  The mixed-precision GCR-DD solver is
illustrated in Algorithm \ref{alg:gcr}.

\begin{algorithm}[htb]
\SetKwData{True}{true}
\SetAlgoLined
\DontPrintSemicolon
\(k=0\)\;
\(r_0 = b - Mx_0\)\;
\(\hat{r}_0 = r_0\)\;
\(\hat{x} = 0\)\;
\While{\(||r_0|| > tol\)}{
\(\hat{p}_k = \hat{K} \hat{r}_k\)\;
\(\hat{z}_k = \hat{M}\hat{p}_k\)\;
\tcp{Orthogonalization}
\For{$i\leftarrow 0$ \KwTo $k-1$}{ 
\(\beta_{i,k} = (\hat{z}_i, \hat{z}_k)\)\;
\(\hat{z}_k = \hat{z}_k - \beta_{i,k}\hat{z}_i\)\;
}
\(\gamma_k = ||\hat{z}_k||\)\;
\(z_k = z_k / \gamma_k\)\;
\(\alpha_k = (\hat{z}_k, \hat{r}_k)\)\;
\(\hat{r}_{k+1} = \hat{r}_k - \alpha_{k}\hat{z}_k\)\;
\(k=k+1\)\;
\BlankLine
\tcp{High precision restart}
\If{\(k=k_{max}\) {\bf or} \(||\hat{r}_k||/||r_0|| < \delta\) {\bf or} \(||\hat{r}_k|| < tol \)} {
\For{$l\leftarrow k-1$ {\bf down} \KwTo \(0\)}{ 
{\bf solve} \(\gamma_l\chi_l + \sum_{i=l+1}^{k-1} \beta_{l,i} \chi_i = \alpha_l\) {\bf for} \(\chi_l\)\;
}
\(\hat{x} = \sum_{i=0}^{k-1} \chi_i \hat{p}_i\)\;
\(x\) = \(x+\hat{x}\)\;
\(r_{0}\) = \(b' - Ax\)\;
\(\hat{r}_0 = r_0\)\;
\(\hat{x} = 0\)\;
\(k=0\)\;
}
}
\label{alg:gcr}
\caption{Mixed-precision GCR-DD solver.  Low-precision fields are
  indicated with a hat (\(\;\hat{}\;\)); e.g., \(x\) and \(\hat{x}\)
  correspond to high- and low-precision respectively.  The
  domain-decomposed preconditioner matrix is denoted by \(K\), and the
  desired solver tolerance is given by \(tol\).  The parameter
  \(k_{max}\) denotes the maximum size of the Krylov subspace.}
\end{algorithm}

\subsection{Improved staggered}

While using a multi-shift CG solver can lead to significant speedups
on traditional architectures, their use is less clear-cut on GPUs:
multi-shift solvers cannot be {\it restarted}, meaning that using a
mixed-precision strategy is not possible; the extra BLAS1-type linear
algebra incurred is extremely bandwidth intensive and so can reduce
performance significantly; multi-shift solvers come with much larger
memory requirements since one has to keep both the \(N\) solution and
direction vectors in memory.

With these restrictions in mind, we have employed a modified
multi-shift solver strategy where we solve Equation
(\ref{eq:multishift}) using a pure single-precision multi-shift CG
solver and then use mixed-precision sequential CG, refining each of
the \(x_i\) solution vectors until the desired tolerance has been
reached.\footnote{Unfortunately, such an algorithm is not amenable to
  the use of half precision since the solutions produced from the
  initial multi-shift solver would be too inaccurate, demanding a
  large degree of correction in the subsequent sequential
  refinements.}  This allows us to perform most of the operations in
single-precision arithmetic while still achieving double-precision
accuracy. Since double precision is not introduced until after the
multi-shift solver is completed, the memory requirements are much
lower than if a pure double-precision multi-shift solver were
employed, allowing the problem to be solved on a smaller number of
GPUs.  When compared to doing just sequential mixed-precision CG, the
sustained performance measured in flops is significantly lower because
of the increased linear algebra; however, the time to solution is
significantly shorter.

\section{Solver performance results} \label{s:Results}
\subsection{Wilson-clover}
Our Wilson-Clover solver benchmarks were run with the QUDA library
being driven by the Chroma \cite{Edwards:2004sx} code. The solves were
performed on a lattice of volume $32^3 \times 256$ lattice sites from
a recent large scale production run, spanning several facilities
including Cray machines at NICS and OLCF, as well as BlueGene/L facilities
at LLNL and a BlueGene/P facility at Argonne National Laboratory
(ANL).  The quark mass used in the generation of the configuration
corresponds to a pion mass of $\simeq$230 MeV in physical units
\cite{Lin:2008pr}.

\begin{figure}[htb]
\begin{center}
\includegraphics[width=3.5in]{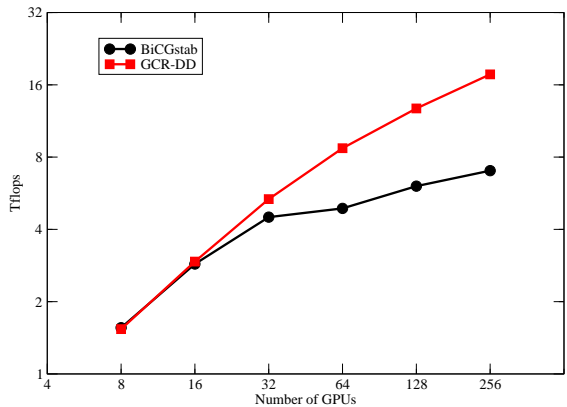}
\end{center}
\caption{\label{fig:clover_flops}Sustained strong-scaling performance in
 Gflops of the Wilson-clover mixed precision BiCGstab and GCR-DD solvers (\(V =
 32^3\times 256\), 10 steps of MR used to evaluate the preconditioner).}
\end{figure}

\begin{figure}[htb]
\begin{center}
\includegraphics[width=3.5in]{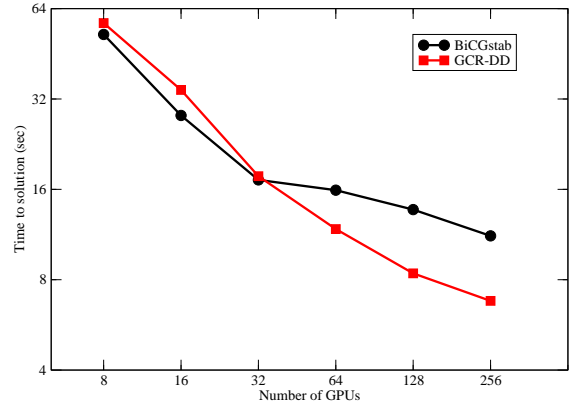}
\end{center}
\caption{\label{fig:clover_time} Sustained strong-scaling time to
  solution in seconds of the Wilson-clover mixed precision BiCGstab and GCR-DD solvers (\(V =
  32^3\times 256\), 10 steps of MR used to evaluate the preconditioner).}
\end{figure}

 In Fig.\ \ref{fig:clover_flops}, we plot the sustained performance in
Tflops of both the BiCGstab and GCR-DD solvers.  For the BiCGstab
solver, we can see that despite the multi-dimensional parallelization,
we are unable to effectively scale past 32 GPUs because of the
increased surface-to-volume ratio. The GCR-DD solver does not suffer
from such problems and scales to 256 GPUs.  As described above, the
raw flop count is not a good metric of actual speed since the the
iteration count is a function of the local block size.  In
Fig.\ \ref{fig:clover_time}, we compare the actual time to solution
between the two solvers.  While at 32 GPUs BiCGstab is a superior
solver, past this point GCR-DD exhibits significantly reduced time to
solution, improving performance over BiCGstab by 1.52x, 1.63x, and
1.64x at 64, 128, and 256 GPUs respectively.  Despite the improvement
in scaling, we see that at 256 GPUs we have reached the limit of this
algorithm.  While we have vastly reduced communication overhead by
switching to GCR-DD, there is still a significant fraction of the
computation that requires full communication.  This causes an Amdahl's
law effect to come into play, which is demonstrated by the fact that
the slope of the slow down for GCR and BiCGstab is identical in moving
from 128 to 256 GPUs.  Additionally, we note that if we perform a
single-GPU run with the same per-GPU volume as considered here for 256
GPUs, performance is almost a factor of two slower than that for a run
corresponding to 16 GPUs.  Presumably this is due to the GPU not being
completely saturated at this small problem size.

\begin{figure}[htb]
\begin{center}
\includegraphics[width=3.5in]{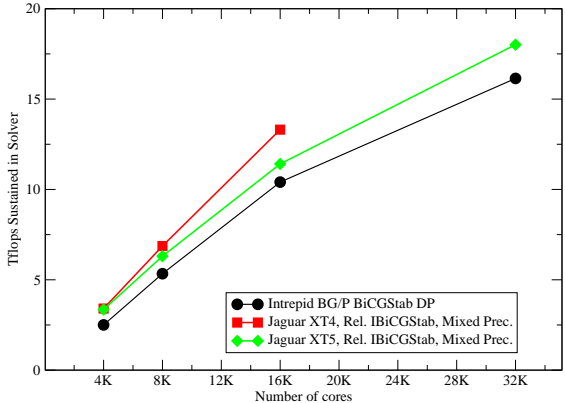}
\end{center}
\caption{\label{fig:arch_compare} Strong-scaling benchmarks on a
  lattice of size $32^3 \times 256$ from Cray XT4 (Jaguar), Cray XT5
  (JaguarPF) and BlueGene/P (Intrepid). Solves were done to
  double-precision accuracy. The Cray solvers used mixed
  (double-single) precision; the BG/P solver used pure double
  precision.}
\end{figure}
  
In terms of raw performance, it can be seen that the GCR-DD solver
achieves greater than 10 Tflops on partitions of 128 GPUs
and above. Thinking more conservatively, one can use the improvement
factors in the time to solution between BiCGStab and GCR to assign an
``effective BiCGStab performance'' number to the GCR solves. On 128 GPUs
GCR performs as if it were BiCGStab running at 9.95 Tflops, whereas on
256 GPUs it is as if it were BiCGStab running at 11.5 Tflops.  To put
the performance results reported here into perspective, in
Fig.\ \ref{fig:arch_compare} we show a strong-scaling benchmark from a
variety of leadership computing systems on a lattice of the same
volume as used here.  Results are shown for the Jaguar Cray XT4
(recently retired) and Jaguar PF Cray XT5 systems at OLCF, as well as the
Intrepid BlueGene/P facility at ANL. The performance range of 10-17
Tflops is attained on partitions of size greater than 16,384 cores on
all these systems. Hence, we believe it is fair to say that the
results obtained in this work are on par with capability-class
systems.

\subsection{Improved staggered}
The results for improved staggered fermions were obtained using the
QUDA library driven by the publicly available MIMD Lattice
Collaboration (MILC) code~\cite{MILC}.  The \(64^3\times 192\) gauge
fields used for this study corresponds to a pion mass of $\simeq$320
MeV in physical units~\cite{Bazavov:2009bb}.

\begin{figure}[htb]
\begin{center}
\includegraphics[width=3.5in]{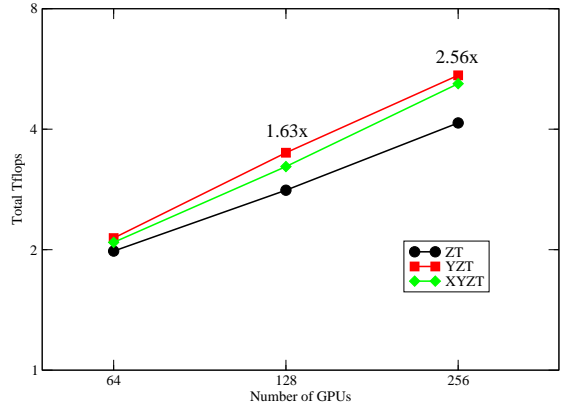}
\end{center}
\caption{\label{fig:stag_cg}Sustained strong-scaling performance in
  Gflops of the asqtad mixed-precision multi-shift solver.  The legend
  labels denote which dimensions are partitioned between GPUs (\(V =
  64^3\times 192\)).}
\end{figure}

In Fig.\ \ref{fig:stag_cg}, we plot the performance of the mixed-precision
multi-shift CG algorithm.  When running the full solver, the minimum
number of GPUs that can accommodate the task is 64.  Reasonable strong
scaling is observed, where we achieve a speed-up of 2.56x in moving
from 64 to 256 GPUs.

With 256 GPUs, we achieve 5.49 Tflops with double-single mixed
precision.  In Ref.\ \cite{Shi:2011ipdps}, we observed an approximately
20\% increase in iteration count for the mixed precision solver,
comparing to the pure double precision one.  To put this in
perspective, the CPU version of MILC running on Kraken, a Cray XT5
system housed at NICS, achieves 942 Gflops with 4096 CPU cores for the
double precision multi-shift solver.  This means one GPU computes
approximately as fast as 74 CPU cores in large-scale runs.

\section{Conclusions} \label{s:Conclusions}
 
Our main result is to demonstrate that by the use of
multi-dimensionsional parallelization and an additive Schwarz
preconditioner, the Wilson-clover solver for lattice QCD can be
successfully strong-scaled to over 100 GPUs. This is a significant
achievement demonstrating that GPU clusters are capable of delivering
in excess of 10 teraflops of performance, a minimal capability
required to apply GPU clusters to the generation of lattice ensembles.
Additionally, multi-dimensional parallelization has enabled the use of
GPUs for asqtad solvers at leading-edge lattice volumes, a feat which
was previously not possible because of the decreased locality of the
asqtad operator.

Clearly, the present use of a simple non-overlapping additive Schwarz
preconditioner is only a first step. It is very likely that more
sophisticated methods with overlapping domains or multiple levels of
Schwarz-type blocking to take advantage of the multiple levels of
memory locality that a GPU cluster offers can be devised to improve
the scaling substantially.  Moreover, we view GPUs and the use of the
Schwarz preconditioner as parts of a larger restructuring of
algorithms and software to address the inevitable future of
heterogeneous architectures with deep memory hierarchies.  We
anticipate that the arsenal of tools needed for the future of lattice
QCD and similarly structured problems (e.g., finite difference
problems, material simulations, etc.) at the exascale will include
domain decomposition, mixed-precision solvers and data
compression/recomputation strategies.

\section{Acknowledgments}
We gratefully acknowledge the use of the Edge GPU cluster at Lawrence
Livermore National Laboratory.  This work was supported in part by NSF
grants OCI-0946441, OCI-1060012, OCI-1060067, and PHY-0555234, as well
as DOE grants DE-FC02-06ER41439, DE-FC02-06ER41440, DE-FC02-06ER41443,
DE-FG02-91ER40661, and DE-FG02-91ER40676.  BJ additionally
acknowledges support under DOE grant DE-AC05-06OR23177, under which
Jefferson Science Associates LLC manages and operates Jefferson Lab.
GS is funded through the Institute for Advanced Computing Applications
and Technologies (IACAT) at the University of Illinois at
Urbana-Champaign.  The U.S. Government retains a non-exclusive,
paid-up, irrevocable, world-wide license to publish or reproduce this
manuscript for U.S. Government purposes.

\bibliographystyle{utcaps}
\bibliography{IEEEabrv}

\end{document}